%% file: main.tex
\PassOptionsToPackage{table}{xcolor}

\documentclass[sigplan]{acmart}
\AtBeginDocument{%
  }

\copyrightyear{2026}
\acmYear{2026}
\setcopyright{cc}
\setcctype{by}
\acmConference[SIGOPS ATC 2026]{The 2026 ACM SIGOPS Annual Technical Conference}{November 15--18, 2026}{Hong Kong, Hong Kong}
\acmBooktitle{The 2026 ACM SIGOPS Annual Technical Conference (SIGOPS ATC 2026), November 15--18, 2026, Hong Kong, Hong Kong}
\acmDOI{10.1145/3828523.3857024}
\acmISBN{979-8-4007-2818-1/2026/11}

\input{c3-latex.tex}

\begin{document}
\pagestyle{plain}
\title{Sharing is caring: Attestable and Trusted Cloud Pipelines out of Distrustful Components}





\author{Amir Al Sadi}
\affiliation{%
  \institution{Imperial College London}
  \country{}
}

\author{Sina Abdollahi}
\affiliation{%
  \institution{Imperial College London}
  \country{}
}

\author{Adrien Ghosn}
\affiliation{%
  \institution{Microsoft Research}
  \country{}
}

\author{Hamed Haddadi}
\affiliation{%
  \institution{University of Oxford}
  \country{}
}

\author{Marios Kogias}
\affiliation{%
  \institution{Imperial College London}
  \country{}
}

\renewcommand{\shortauthors}{Al Sadi et al.}

\input{sections/abstract.tex}


\maketitle

\section{Introduction}
\input{sections/introduction_asplos.tex}

\section{Background \& Motivation}
\input{sections/background_motivation.tex}
\section{Design}
\input{sections/design.tex}

\section{Implementation}
\input{sections/rmm_impl.tex}

\section{Infrastructure}
\input{sections/infrastructure.tex}

\section{Evaluation}
\input{sections/evaluation_asplos.tex}

\section{Related Work}
\input{sections/related_work_asplos.tex}

\section{Conclusion}
\input{sections/conclusion.tex}

\begin{acks}
\input{sections/acknowledgments_final.tex}
\end{acks}

\bibliographystyle{ACM-Reference-Format}
\bibliography{gen-abbrev,dblp, misc}

\clearpage
\appendix
\input{sections/policy_appendix.tex}
\input{sections/cca_bg_appendix.tex}

\appendix

\end{document}

%% file: c3-latex.tex
\newif\ifcomments

\usepackage{xspace}                         
\usepackage{caption}
\usepackage{subcaption}
\usepackage{comment}                        
\usepackage{rotating}
\usepackage{graphicx}
\usepackage{multirow}
\usepackage{cleveref}
\usepackage{listings,lipsum}
\usepackage{enumitem}
\usepackage{listings}
\usepackage{xcolor}
  \usepackage[table]{xcolor}

\lstdefinelanguage{json}{
  basicstyle=\ttfamily\small,
  numbers=left,
  numberstyle=\tiny,
  stepnumber=1,
  numbersep=5pt,
  showstringspaces=false,
  breaklines=true,
  frame=lines,
  morestring=[b]",
  stringstyle=\color{blue},
  literate=
   *{0}{{{\color{black}0}}}{1}
    {1}{{{\color{black}1}}}{1}
    {2}{{{\color{black}2}}}{1}
    {3}{{{\color{black}3}}}{1}
    {4}{{{\color{black}4}}}{1}
    {5}{{{\color{black}5}}}{1}
    {6}{{{\color{black}6}}}{1}
    {7}{{{\color{black}7}}}{1}
    {8}{{{\color{black}8}}}{1}
    {9}{{{\color{black}9}}}{1}
}

\usepackage[skip=0pt]{caption}
  \usepackage{ifthen}

\makeatletter
\@ifclassloaded{sigplanconf}{
}{
}
\makeatletter

\makeatletter
\let\originalunderscore\_
\renewcommand{\_}{%
  \originalunderscore
  \ifmmode\else
    \ifx\f@family\ttdefault
      \allowbreak
    \fi
  \fi
}
\makeatother

\newcommand{\printlayout}{{\bf The textwidth \the\textwidth; the columnwidth is \the\columnwidth}}

\newcommand{\eg}{\textit{e.g.,}\xspace}
\newcommand{\ie}{\textit{i.e.,}\xspace}

\newcommand{\myparagraph}[1]{\vspace{0.2em}\noindent {\bf #1.}}

\usepackage[english]{babel}
\addto\extrasenglish{%
  \def\chapterautorefname{\S\@gobble}%
  \def\sectionautorefname{\S\@gobble}%
  \def\subsectionautorefname{\S\@gobble}%
  \def\subsubsectionautorefname{\S\@gobble}%

  }

\newcommand{\system}{\textsc{Mica}\xspace}
\newcommand{\systems}{\textsc{Mica's}\xspace}
\newcommand{\arch}{Arm\xspace}
\newcommand{\realm}[1]{\emph{R#1}}

%% file: sections/abstract.tex
\begin{abstract}
The trust model underlying today's Trusted Execution Environments (TEEs) is inadequate for modern confidential workloads.
TEEs protect private memory and attest initial code and configuration, but they neither constrain nor attest the shared channels through which data enters, leaves, or flows between TEEs or with the untrusted host.
As a result, confidentiality remains coupled to trust in the functional correctness of all in-TEE code that may access those channels.
This trust assumption is increasingly untenable for agentic applications that generate code at runtime and for cloud pipelines composed of mutually distrustful components.
A single misuse of a shared channel is enough to leak confidential data and compromise the entire pipeline.

We present \system{}, a confidential-computing architecture firmware extension that decouples confidentiality from trust in in-TEE code.
\system{} lets tenants explicitly define, restrict, and attest channel constraints between TEEs and with the untrusted host, including transitive constraints that span an entire pipeline.
It further extends attestation from individual TEEs to groups of communicating TEEs, allowing external parties to verify the confidentiality properties induced by those constraints across a whole pipeline.
We implement \system{} on Arm CCA by extending the Realm Management Monitor with a small policy language, requiring no hardware changes and only minor changes to QEMU and KVM.

Our evaluation shows that \system{} supports modern confidential agentic applications and pipeline computations without requiring trust in the functional correctness of in-TEE code.
Our implementation increases the TCB by only 7\% and incurs infrequent setup costs: 2--4\% boot-time overhead, approximately 1.5\,s for policy upload, and 246--598\,ms for group attestation.
Once established, Realm communication is direct, uses protected memory, and requires no \system{} intervention or software encryption leading to up to $14.5\times$ lower cross-TEE message latency than encrypted communication over untrusted memory.

\end{abstract}

%% file: sections/introduction_asplos.tex

Confidential computing protects data during a computation, providing the necessary complement in the cloud to existing protections for data at rest and in transit.
The ecosystem is maturing rapidly: hardware vendors ship Trusted Execution Environments (TEEs)~\cite{tdx, amd, cca}, cloud providers offer confidential-computing services built on them~\cite{google-cc,azure-cc,aws-cc}, and a growing body of academic work extends them further~\cite{verismo, veil, erebor, guarantee}.
The trust model underpinning TEEs, however, was conceived for software that is static, auditable, fully measured at deployment, and functionally trusted to handle confidential data without leakage --- assumptions that no longer hold for emerging workloads, particularly agentic applications~\cite{agentee} that generate and execute code at runtime.

Today's TEEs adopt a myopic view of confidentiality, providing only half of what is needed.
The first half is covered by hardware: it protects a TEE's \emph{exclusive} resources, above all its private memory, from higher-privilege software, and \emph{attests} a launch-time measurement of the TEE's initial code and configuration~\cite{cca,tdx,amd}.
The other half is not: a TEE's confidentiality also depends on its \emph{shared} channels, as shown in \autoref{fig:positioning}.
These include shared memory, communication buffers, and control transfers to untrusted software.
Existing TEE architectures neither constrain what may flow across these channels nor attest any policy that bounds them.

This blind spot forces a fundamental \emph{conflation of confidentiality and trust}: the only way to guarantee confidentiality is to trust all TEE-resident code that accesses shared channels to prevent data leakage.
In this model, every such component becomes part of the Trusted Computing Base (TCB) for confidentiality.
For this model to work, the TCB must be stable enough for tenants to know which measurement to expect at attestation time, small enough to audit, and functionally trusted to use shared channels without leakage.

That trust is brittle: shared channels are broad, unconstrained, and neither measured nor bounded by attestation.
Code that can access them is rarely stable: modern software stacks evolve through frequent updates and dependencies that may change independently.
Nor is it small enough to audit: these stacks combine complex low-level systems code (\eg virtio~\cite{virtio} drivers) with many libraries and elaborate dependency patterns, making comprehensive audits difficult.
Most importantly, correctness remains a functional assumption, so a single bug or misuse in code that accesses shared channels is enough to break confidentiality.

\input{images/mica_positioning}

To make matters worse, modern workloads may not even expose the relevant code at attestation time.
Agentic applications make this problem structural: the logic that handles confidential data may be generated or selected at runtime in response to LLM outputs and external tool calls~\cite{aws-agents, azure-agents}, rather than being static, auditable, and fully present at deployment.
The problem extends beyond agentic systems: cloud workloads increasingly run as CVM pipelines operated by separate parties, with proprietary code, asymmetric trust, and limited willingness or ability to audit one another.
In short, today's TEEs offer confidentiality only at the price of a trust assumption modern workloads cannot pay.

Prior systems try to tackle this challenge by confining untrusted code from within the TEE,  but either constrain that code too tightly for modern workloads or provide only per-TEE guarantees rather than pipeline-wide ones.
Ryoan~\cite{ryoan} demonstrated strong sandboxing on SGX, but requires recompiling untrusted code into a restricted execution model, is limited to application-level confinement, and requires unanimous trust in the sandboxing code --- a poor fit for agentic workloads that need rich environments and dynamically generate or execute code.
Paravisor-based designs~\cite{veil, verismo, erebor} use intra-TEE privilege layers (\eg VMPLs in AMD SEV-SNP~\cite{amd}) to mediate an unmodified guest's interactions with the host. However, they observe and constrain only their own CVM, not the broader pipeline.
Pipeline participants must standardize on the same stack inside each CVM, while guarantees remain implicit in the mediator’s code rather than explicit, attestable constraints across the pipeline.
The missing step is to stop making TEE-resident software the basis for confidentiality.


Tenants need to \emph{decouple confidentiality from trust}: data should remain confidential across computation pipelines even when intermediary components run code tenants do not trust.
Trust should anchor only at the pipeline's ingress and egress, where data enters and leaves and must be made explicit, observable, and attestable.

To achieve this, we take a different approach: instead of constraining code inside each TEE, we measure and constrain the channels between TEEs and with the host.
By doing so, we target the blind spot of today's TEEs: confidentiality follows from allowed data flows, not from the correctness of in-TEE software.
Under these principles, confidentiality requires two conditions:
(1)~exclusive memory remains inaccessible to external entities --- a property already provided by existing TEEs --- and
(2)~all channels between components are explicitly defined, restricted, and attested.
This is a modern take on Lampson's classic \textit{Confinement Problem}~\cite{confinement}: hardware--software co-design is required to prevent information leaks across component boundaries.

We realize this approach in \system{}\footnote{https://github.com/c3infer/c3infer/}: a confidential-computing architecture extension that decouples confidentiality from trust in code.
\system{} strenghtens confidentiality across pipeline computations without requiring tenants to trust the code running in individual TEEs.
It does so by making shared channels explicit, attested, and constrained by tenant-defined policies: unauthorized channels are eliminated, and TEEs can communicate only when their policies are compatible.
These channels include memory shared with other TEEs, memory shared with the untrusted host, and control-flow transitions to the host that can serve as covert channels, but do not include timing or microarchitectural channels, which remain outside our threat model.
Confidentiality then follows from constrained data flow across authorized channels, rather than trust in the code that handles it.

\system{} policies are transitive: they constrain not only a TEE's channels to its direct \emph{peers}, but also how those peers may further communicate.
This lets tenants use a policy language to express constraints over an entire \system{} \emph{group} of communicating TEEs --- even without a complete view of the pipeline --- rather than only over pairwise links.
\system{} attests each group and its policies, allowing external parties to verify the resulting end-to-end confidentiality properties.

We implement \system{} on \arch{} CCA by extending the Realm Management Monitor (RMM), adding 7\% to the TCB, requiring no hardware changes, and only minor modifications to QEMU~\cite{qemu} and KVM~\cite{kvm}.
Our prototype builds on CAEC's protected memory channels~\cite{sinas-arxiv} for direct CVM-to-CVM channels, and extends them with \system{} policies over channels between TEEs and with the untrusted host.
Our prototype is limited to \arch{} CCA; we leave validating the same principles on platforms with similar firmware and memory-sharing primitives, such as Intel TDX~\cite{tdx}, to future work.
We also use Alloy to specify \system's policy language and model check reachability between Realms for different configurations.

Our evaluation demonstrates that \system{} supports use cases ranging from unidirectional video processing pipelines and guarded LLMs to trusted network proxies for untrusted workloads, without trusting the functional correctness of every in-TEE service for confidentiality.
With strict policies that fully describe every component in a pipeline, \system{} can ensure that an entire pipeline is booted, configured, and attested before secrets are provisioned.
With relaxed policies, it can instead mediate access to resources through a trusted component, such as an attested network proxy, for untrusted workloads that are not known in advance.
In both cases, \system{} statically enforces the permitted communication paths, while trusted components can dynamically update the higher-level policies enforced along those paths, such as network filters or agent tool authorizations.

On our OpenCCA~\cite{opencca} implementation, \system{} strengthens CAEC's security guarantees by removing its mutual-trust assumptions without adding communication overhead, while incurring only modest boot (2--4\%), policy setup ($\sim$1.5\,s), and group-attestation (246\,ms/598\,ms for two/three Realms) costs.

This paper makes the following contributions:
\begin{itemize}
\item The design of \system{}, which decouples confidentiality from trust in in-TEE code through three key shifts: making shared channels explicit, replacing implicit policies with attestable constraints, and deriving confidentiality from constrained data flow rather than code correctness;
\item A policy language that allows tenants to explicitly define, restrict, and attest pipeline-wide channel constraints between TEEs and with the untrusted platform, along with Alloy specification of the language;
\item An implementation of \system{} on \arch{} CCA, demonstrating feasibility with a modest TCB increase and no hardware modifications.
\end{itemize}

%% file: images/mica_positioning.tex
\begin{figure}[t]
  \centering
  \includegraphics[width=0.55\linewidth]{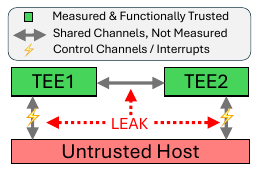}
  \caption{Normal CC attests each TEE image at boot, but not the channels through which TEE data may later leak.}
  \label{fig:positioning}
  \vspace{-10pt}
\end{figure}

%% file: sections/background_motivation.tex
The conflation of confidentiality and trust in today’s TEEs fundamentally limits how confidential pipelines can be built out of mutually distrustful components.
We first study how TEEs communicate and the security and performance challenges these interfaces introduce.
We then examine prior approaches to building confidential pipelines and show where they fall short in managing trust across components.

\subsection{TEE Interfaces and Communication}

\input{sections/saved_background_tee.tex}

\subsection{Approaches to Confidential Pipelines}
\label{sec:motivation}
\input{images/strawmans_slim}

We study confidentiality challenges in building a multi-component confidential-cloud deployment.
Consider an agentic application where a user interacts with an LLM via an agent running on the same machine.
The agent and LLM are proprietary, developed by different organizations, and user prompts are sensitive.
Trust relationships are asymmetric: the user trusts the agent provider but not the model provider; the agent and model providers do not trust each other; and none of the parties trust the cloud infrastructure.

We examine a series of increasingly sophisticated deployment strategies, highlighting the confidentiality, performance, and trust challenges each addresses.

\myparagraph{Naive multi-CVM deployment -- \autoref{fig:strawman-designs} (a)}
One straightforward approach is to run the agent and LLM in separate CVMs on the same host, communicating over the local virtual network.
This ensures isolation from the cloud provider and from each other, but confidentiality depends on trusting both CVMs’ entire software stacks -- including the OS and user-space programs -- to correctly handle sensitive data.
All communication passes through hypervisor-visible channels, requiring software encryption and decryption, and each component must trust the other to not leak data via the host.
This approach highlights the full set of problems: large TCB, cross-tenant trust requirements, and costly overhead of cross-TEE communications due to encryption and protection against the hypervisor's access to the channel.

\myparagraph{CVMs with intra-VM sandboxes -- \autoref{fig:strawman-designs} (b)}
A first improvement is to reduce the TCB by introducing sandboxes inside each CVM to mediate information flow.
Sandboxes may run in user-space (\eg NaCl~\cite{nacl}, Ryoan~\cite{ryoan}) or rely on OS-level mechanisms (\eg gVisor~\cite{gvisor}).
This removes most application code from the TCB, but the sandbox and OSes, commonly maintained by different companies, must still be trusted.
The rest of the pitfalls remain the same.

\myparagraph{Single CVM with multiple sandboxes -- \autoref{fig:strawman-designs} (c)}
Next, collocating all components within a single CVM and isolating them using software sandboxes further reduces the TCB: only the OS and the sandbox mechanism (\eg containers or gVisor) must be trusted.
Communication occurs entirely within the CVM, avoiding hypervisor-visible channels and the associated encryption overhead.
However, tenants must coordinate to attest both the CVM and the sandbox before provisioning secrets, which limits independent updates.
Moreover, the OS remains part of the TCB to mediate interactions with the host.
CHANCEL~\cite{chancel} follows this approach within an SGX enclave, using compiler-instrumented SFI and a trusted runtime to isolate per-client data within a potentially adversarial service.

\myparagraph{Monitor/Paravisor for intra-CVM isolation -- \autoref{fig:strawman-designs} (d)}
Hardware mechanisms, such as AMD VMPLs~\cite{amd} or Intel Trust Domains~\cite{tdx}, provide intra-CVM compartmentalization.
A trusted monitor or paravisor running at the highest privilege level enforces isolation by (1) shielding CVM content and mediating interactions with the host, and (2) isolating components within the CVM.
Compartments can securely share memory under monitor mediation, eliminating host-exposed channels and removing the need to trust the OS or an application-level sandbox. 
However, the number of compartments is often fixed and small (\eg $1+3$ in TDX), the monitor itself must be fully trusted, and secondary attestation for individual components is required.
Existing solutions enforce fixed policies tied to specific trust models: Verismo~\cite{verismo} shields trusted modules, Veil~\cite{veil} implements nested enclaves, and OpenHCL~\cite{openhcl} provides compatibility for unenlightened CVM guests.
As a result, confidentiality remains dependent on the correctness of the trusted code.



Across all prior approaches, achieving confidentiality comes at the cost of one or more trade-offs: high communication overhead, requiring all components to be trusted, or forcing components to be managed together rather than independently.
\system{} breaks this trade-off by giving tenants explicit, attestable control over all communication paths in the pipeline.
This enables components to remain independently managed while still providing strong, end-to-end confidentiality guarantees for the data they process.

%% file: sections/saved_background_tee.tex
TEEs, regardless of hardware vendor, adopt a binary memory model: memory is either trusted or untrusted.
Trusted memory is accessible only by the TEE, while untrusted memory is fully managed by the hypervisor or host and can potentially be shared with other TEEs.
Any data placed in untrusted memory is therefore exposed to the host, which controls both its allocation and contents.
Shared memory provides the canonical mechanism for a TEE to receive input or produce output accessible to the outside world, forming the basis of communication and interaction between the TEE and the host, hence with other TEEs externally, as well.

For confidential virtual machines (CVMs), the most common TEE paradigm, shared memory is essential for supporting paravirtualized devices such as storage and networking, hypercall-based services, and emulating elements of the boot process (\eg ACPI tables) to maintain compatibility with existing software and enable \textit{lift-and-shift} deployment~\cite{gramine-tdx, teeio}.

Relying on hypervisor-accessible shared memory introduces two fundamental challenges.
The first is performance.
Communication through shared memory requires copying data between trusted and untrusted memory.
To preserve confidentiality, TEEs must additionally encrypt data before writing and decrypt it after reading, adding further overhead.
Prior work~\cite{bifrost,cvms-explained} shows that this combined overhead can consume up to 50\% of CPU cycles.
A similar performance cost applies to TEEs collocated on the same host that exchange sensitive data via shared memory.

The second challenge is security.
Interfaces such as virtIO~\cite{virtio} or ACPI~\cite{acpi} were not designed for a confidential setting, where the hypervisor or host may be untrusted.
Hardening these interfaces is either extremely challenging and error-prone~\cite{teeio}, or neglected entirely~\cite{badaml}.
Consequently, attacks exploiting the boundary between a TEE and the host are well documented~\cite{control-chan-atks,pigeonhole-sgx}.
While prior work has progressively hardened some interfaces, this approach ultimately shifts the burden onto TEE code and its functional correctness, which is difficult to guarantee in practice.
TEE-to-TEE communication over shared memory remains exposed to the host and requires proper hardening of both endpoints to protect against attacks, such as replayed messages.

Several approaches attempt to address these limitations by enabling confidential shared memory among TEEs collocated on the same host.
Systems such as Elasticlave~\cite{elasticlave}, Plug-In Enclaves~\cite{plug-in-enclave}, and Cerberus~\cite{cerberus} allow TEEs to access the same physical memory without exposing it to the untrusted hypervisor.
These approaches reduce the overhead of copying and encryption and provide a starting point for \system, yet they are not widely available across TEE architectures.
Similarly, emerging confidential I/O devices that rely on mutual attestation and encrypted PCIe communication allow direct TEE-to-device access, though these remain limited primarily to certain GPUs~\cite{volos-conf-gpus}.

%% file: images/strawmans_slim.tex
\begin{figure}
    \centering
    \includegraphics[width=0.8\linewidth]{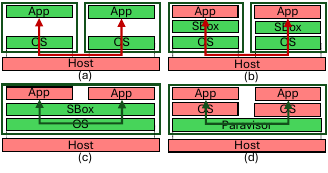}
    \caption{Prior confidential-pipeline designs. Red denotes untrusted components or channels; green denotes trusted components or channels. Dotted lines mark interfaces exposed to the untrusted host.}
    \label{fig:strawman-designs}
    \vspace{-10pt}
\end{figure}

%% file: sections/design.tex



\label{sec:design}




We present \system, a system which generalizes confidential computing from protecting a single trusted component to securing end-to-end application pipelines composed of mutually untrusted components.
In \system, each component executes in its own isolated TEE instance and interacts with others only through controlled and attested communication channels.
Unlike prior TEE solutions, \system does not require trusting the functional correctness of any component.
Instead, confidentiality is enforced by construction through these controlled and attested inter-TEE communication channels, allowing security properties to extend across an entire pipeline rather than being confined to a single TEE.

\system achieves this with a software extension to TEE architectures that supports user-defined policies over:
(i) unprotected memory;
(ii) shared protected memory between TEEs; and
(iii) known covert channels.
Finally, a \emph{group} attestation scheme captures these policies across all communicating TEEs, allowing to reason about information sharing and leakage across an entire pipeline or directed graph of interacting components.

\subsection{Threat Model}
\label{sec:design:td}

\system targets a multi-component application deployed in the cloud.
The application consists of multiple components from different vendors.
Component vendors do not trust one another nor the cloud provider.


The application processes sensitive data that must remain confidential throughout the processing pipeline.
Client data must not be disclosed to the cloud provider or exfiltrated by individual components.
Although the client assumes that each component performs its advertised functionality, it does not trust components to preserve confidentiality and treats them as adversarial with respect to data leakage, including through collusion with the cloud provider.

The client and software vendors trust the correctness of the underlying TEE hardware and firmware, which together form the TCB.
The client does not, in general, trust the functional correctness of components' code, even when it executes inside a TEE.
However, components directly controlled by the client may be considered trusted and part of the TCB.

Denial-of-service attacks are out of scope.
\system{}'s extended RMM, which is part of the TCB and covered by attestation, limits data leakage through explicit communication paths by blocking host transitions or scrubbing their arguments, and provides extensible mechanisms for best-effort mitigation of covert channels through these paths.
It does not eliminate timing or microarchitectural channels, which remain outside our threat model; complementary defenses can be used alongside \system{} to address them.

\subsection{Design Overview}
\label{sec:design:high-level}

\input{images/mica_overview}

In \system, each component executes in its own TEE, co-located on the same host.
Communication between components is governed by explicit, system-enforced \emph{policies}.
Policies precisely define how components share memory and constrain how information flows across component boundaries.


A key design principle of \system is to make memory sharing -- and hence communication between components -- explicit and attestable.
In existing TEE architectures, sharing with untrusted entities such as the hypervisor is implicit and unchecked, which forces confidentiality to rely on trusting TEE code not to leak sensitive data.
\system breaks this assumption by requiring tenants to explicitly specify, via policies, how a TEE may share memory with other TEEs or with the host.
By enforcing explicit sharing, \system decouples confidentiality guarantees from trust in component behavior.


However, explicit sharing alone is insufficient.
Although it restricts the behavior of an individual TEE, it does not control how data may propagate once shared with other components.
To address this limitation, \system takes a second, \emph{inductive} step: policies do not only describe how a component may share memory, but also constrain the communication behavior of the components it connects to.
Concretely, a component’s policy specifies properties that peer components must satisfy.
By enforcing these constraints at connection time, \system allows components to bound the transitive flow of data beyond their own execution context.

Even with explicit memory sharing and transitive bounds on information flow, components may still establish \emph{covert channels} to communicate.
In particular, a component can repurpose existing signaling mechanisms --- most notably interactions with the untrusted host --- to leak confidential information.
For example, it may encode data in the frequency of exceptions, in the timing of transitions, or in the parameters passed during host interactions.
To address this, \system{}'s policies cover such interactions explicitly.
They specify which signaling mechanisms a component is permitted to use, and under what conditions.

Finally, \system elevates attestation from individual TEEs to entire application pipelines.
By attesting a group of communicating TEEs together with their declared policies, \system allows external users to verify that the entire deployment satisfies the desired end-to-end confidentiality properties, rather than reasoning about individual components.

To achieve the above goals, \system extends existing TEE architectures and makes the following three assumptions about the underlying TEE implementation.
First, the TEE offers a way to extend its functionality without requiring hardware changes.
Second, the extensible firmware is able to interpose and filter the transitions and interaction between TEEs and the untrusted host.
Third, the TEE implementation allows for confidential shared memory~\cite{sinas-arxiv}, \ie shared memory between different TEEs that is not accessible by the untrusted host.
These assumptions are compatible with TEEs across different hardware platforms, such as M-mode in RISC-V~\cite{elasticlave} and the RMM in \arch{} CCA~\cite{cca}.
Thus, we describe the rest of the design in an architecture-agnostic way and we specialise our implementation (\S\ref{sec:rmm_impl}) on \arch{} CCA.

Figure~\ref{fig:mica-overview} gives the high-level overview of \system for an agentic pipeline.
Different collaborating TEEs form a group (grey blob).
Each TEE comes with a policy file and uploads this policy to \system.
The Agent and Network TEEs have already uploaded policies $P_A$ and $P_N$.
For a new TEE to join a group (LLM in the scenario described in the figure), \system needs to validate if the new policy is compatible with the rest of the policies in the group.
If validation succeeds, \system admits the LLM to the group and activates its authorized channels; otherwise, the new peer is rejected.
The group attestation will include the TEE measurements, the existing communication channels, and the provided policies.
In this deployment only the Network TEE is part of the TCB and is the only one that can directly communicate with the untrusted host.

\subsection{\system Policies}
\label{sec:design:policy}

While seemingly straightforward, designing the right support for policies over communications poses two key challenges.
First, policies must balance flexibility and precision: they need to provide strong guarantees, such as preventing leakage of sensitive information, while allowing components to be managed and updated independently.
Second, enforcing restrictions on interactions with untrusted code is inherently difficult. TEEs were not originally designed for this purpose and often rely on the untrusted hypervisor during critical phases of their lifecycle, such as boot.
We address both challenges by defining semantics that make minimal assumptions about communicating components and adopting a semi-dynamic policy model: policies can be supplied at runtime, after boot, at which point they begin to be enforced and can no longer be changed.


\myparagraph{Terminology}
In \system{}, a component's \emph{policy configuration} defines its communication \emph{channels} and the \emph{peers} it connects to.
Channels span two types of media: memory and control-flow transitions.
Memory channels enable communication between TEEs or between a TEE and the host via shared memory, while control-flow channels occur only between a TEE and the host.
A channel is \emph{protected} if it is accessible only to TEEs, and \emph{unprotected} if the host can access it.
Channels are \emph{active} when shared by two or more peers, and \emph{inactive} when accessible to only a single component.

For each peer, a component's policy can specify authentication requirements and impose constraints on the peer’s own channels. These transitive constraints implement \system{}’s inductive step, allowing a component to control not just its direct communications but also how information may propagate through its peers.

Peers permitted to interact with untrusted code or access unprotected memory are called \emph{gateways}.
At run-time, the set of components connected by active channels -- directly or transitively -- forms a \emph{group}.

\myparagraph{Component policy lifecycle}
Any running TEE may supply a policy to \system{} to govern its communication.
Until a policy is provided, the TEE executes without restrictions, allowing it, \eg to boot with the help of an untrusted hypervisor.
A TEE supplies its policy configuration via a dedicated \system{} call.
The policy takes effect immediately, is enforced by \system{}, and cannot be modified.
If the supplied policy cannot be satisfied or conflicts with the policies of other peers in the group, \system{} terminates the component.
Otherwise, the component joins the group, is allowed access to the protected channels, and its interactions with the host are limited to its restricted control-flow and unprotected memory channels.



\myparagraph{Policy configuration structure}
A policy configuration contains two types of entries: (1) peers, and (2) channels (see Appendix~\ref{sec:appendix:policy}, Fig.~\ref{fig:policy} for an example).


Peers include the component itself, indicated by the keyword \texttt{Self}, as well as other components in the group from the supplier’s perspective.
For each peer, the policy specifies a local ID for referencing it elsewhere in the configuration, an optional hash of the peer’s initial attestation report, and a set of expected properties.
The properties consist of two booleans: \emph{is\_gateway} indicates whether the peer may access unprotected memory,  and \emph{strict} to indicate whether it may maintain channels not explicitly listed in the current policy.
The strict boolean is important for decoupling peer management: it allows components that do not require strict confidentiality to avoid tracking a peer’s channels by setting it to \texttt{false}.
For peers with $strict = true$, all their channels and peers must be captured in this policy configuration.
Channels, whether based on memory or covering control-flow, appear as a list.
Policies are measured separately by \system{} rather than included in the initial TEE measurements, avoiding circular dependencies with policy-based authentication between components in the confidential pipeline.


\myparagraph{Memory channels}
For a memory channel, the policy specifies the following fields: \emph{size}, \emph{type}, and a set of memory \emph{mappings}.
The \emph{size} field specifies the size of the shared region; for confidential virtual machines, this corresponds to the contiguous size of the mapping in guest physical address (GPA) space.
The \emph{type} is either \texttt{protected} or \texttt{unprotected} and indicates if the physical memory backing the channel must be confidential, \ie not accessible by the host, or not.

The \emph{mappings} field is central to a channel specification, defining how peers are connected.
It contains one entry per peer with shared-memory access, identified by the peer’s local ID or \texttt{SELF}. 
Each entry may optionally specify the guest physical address where the channel is expected to be mapped, along with maximum access rights (read, write, execute).
Although executable shared memory is uncommon, supporting it adds no extra complexity and lets one peer supply code to another without costly private-memory copies.

To provide flexibility, \system{} also supports the special identifier \texttt{any}, which represents components not explicitly listed as peers.
A channel mapping can use \texttt{any} to allow additional components, not described in the policy, to access the channel.
An \texttt{any} mapping includes the same fields as above and an additional counter that limits the number of components that can use this entry to connect to the channel, with negative values indicating no limit.
This mechanism allows components to act as proxies, granting arbitrary peers access to specific services, for example, devices.

\myparagraph{Control-flow channels}
Control-flow channels specify policies to restrict information carried when transitioning to the host and mitigate their use as covert channels.
A control-flow channel policy has three fields: \emph{owner}, \emph{type}, \emph{range}, and \emph{policy}.
The owner field is the local ID for the peer to which the policy applies.
The type identifies the transition trigger and can take the values of \texttt{call} indicating an explicit call from the TEE to the host, or \texttt{exception} indicating an implicit transition to the host based on the TEE execution, \eg a page fault.
The range defines a non-empty set of identifiers describing the specific transition type.

The policy field specifies how \system{} handles such control-flow events and can take three values: \texttt{allow}, \texttt{scrub}, or \texttt{block}.
\texttt{allow} enables such transitions to the host along with the information transfers as described by the ISA.
The \texttt{scrub} policy instructs \system to obfuscate the related information before transitioning to the host.
The obfuscation logic is specific to the transition and implemented in \system{}.
Conceptually, this policy filters and sanitizes interrupt-related information before control is transferred to the untrusted host.
It can further be augmented to consider the rate at which such events are trigger and introduce noise in a best-effort attempt to mitigate covert channels based on timing.
Finally, \texttt{block} drops the event completely and is the default when no policy is explicitly supplied for the event.
To prevent a TEE's policy from hogging a core, \system{} can selectively force a platform specific event (\eg a particular IPI on x86 or a secure timer on \arch{}) to be scrubbed rather than blocked.

\myparagraph{Default behaviour}
\system policies define explicit memory sharing and control flow transitions.
\system enforces the invariant that any memory available to the TEE and not described by the policy file is protected (\ie confidential) and only accessible by this TEE.
\system{} further ensures that any protected memory used for a channel is mapped only once in the TEE's address space.
This is necessary to avoid attacks that would alias the same physical memory at another guest physical address space with different access rights.

\subsection{Policy Validation \& Enforcement}
\label{sec:design:pve}

\system{} extends TEE operations and is designed to be backward compatible with existing technologies.
\system{} enables TEEs that do not use its extensions to operate normally.
To this end, our design requires a TEE to explicitly join a group by uploading its policy.
At this point, \system{} validates the policy, ensuring it can be satisfied, before applying restrictions to the domain and enabling channels. 
Failure to validate the supplied policies results in \system{} terminating the TEE.

\myparagraph{Before policy upload}
Prior to uploading a policy to \system{}, a TEE operates normally.
It has unsupervised access to unprotected memory as allowed by the untrusted host, does not share any protected memory with other TEEs on the system, and can request platform-provided attestations.
Note that for TEE technologies that support dynamic memory management, \ie adding protected memory at run-time via page-faults or collaborative protocols with the host, we recommend pre-populating the TEE's entire address space \emph{before} uploading the policy configuration as it might restrict the control-flow transitions necessary to populate them.
For memory backing protected memory channels, \system{} requires the host to mark the corresponding regions explicitly as protected shared and to pre-supply the backing physical pages. 
It however blocks the TEE's access to these regions until the policy upload.
Internally, \system{} maps protected-shared physical addresses to the TEEs containing them.

\myparagraph{Upload and validation}
A policy upload triggers \system{}'s validation of the policy.
The TEE enters a locked-down state that prevents any modification to its memory mappings by the host.
\system{} parses the policy and inspects the TEE's address space to validate whether these can be satisfied.
First, \system{} goes through all the TEE's unprotected memory and unmaps any access that is not explicitly covered by the policies.
Second, \system{} ensures that there is a one-to-one correspondence between mappings marked by the host as protected shared and the protected channels described in the policy.
It further ensures the size of these regions is correct and that the same physical page is not used more than once.
At this stage, \system{} is ready to validate it against other peers policies.

\system{} validates policies by performing a graph traversal in which nodes represent peers and edges represent memory channels defined by the policies.
Starting from the peer that uploads its policy, \system{} initializes a work queue with all channels described in that policy.
For each channel dequeued, and for each of its mappings, \system{} resolves the peer’s local identifier to a globally unique TEE identifier maintained by \system{}.
This resolution is performed either by matching the peer’s supplied attestation hash, if present, or by consulting the mapping between shared protected physical pages and the TEEs that map them and already uploaded their policies.
If \system{} cannot unambiguously resolve peer identifiers, policy validation fails and the initiating TEE is terminated.

Once a peer is successfully identified, \system{} caches the local-to-global mapping and checks that the policies on both sides are mutually compatible.
This includes validating peer properties as well as channel attributes such as access rights and size.
The channel's mapping is then marked as visited, and all channels described in the peer’s policy are added to the work queue.
Already visited channel mappings are skipped when encountered again.

When the traversal completes successfully, meaning all reachable channels have been validated, \system{} enables the initiating TEE’s access to the channels according to its policy.
Note that active channels are the ones for which the physical address to peer map contains at least two peers with uploaded (and validated) policies.

\subsection{Attestation}
\label{sec:design:attestation}

\system{} extends standard TEE attestation by incorporating the component’s supplied policy and the current state of its channels.
Crucially, for backward compatibility, \system{} leaves the default TEE attestation unchanged, ensuring that components remain interoperable with non-\system{} services.
This way vendors can independently attest that their components run inside a TEE implemented by a platform they trust.
This is possible before a component installs its policy and removes unprotected mappings, or anytime for gateways.

\system{} provides an extended attestation that produces a single report covering an entire group.
To construct this report, \system{} performs a graph traversal over active channels, visiting all TEEs reachable from the initiating peer.
For each peer in the group, the attestation includes its standard platform attestation and its policy configuration, with channels marked as active or inactive.
These artifacts are included in plaintext, alongside a cryptographic hash of their contents that is signed by \system{}.
The exact signature format and verification procedure are platform- and implementation-specific and are implemented on top of the platform’s existing mechanisms and roots-of-trust.
Note that any peer within the group  obtains the same group attestation but that only gateways are capable of sending the attestation to a remote client.

%% file: images/mica_overview.tex
\begin{figure}[t]
  \centering
  \includegraphics[width=0.75\linewidth]{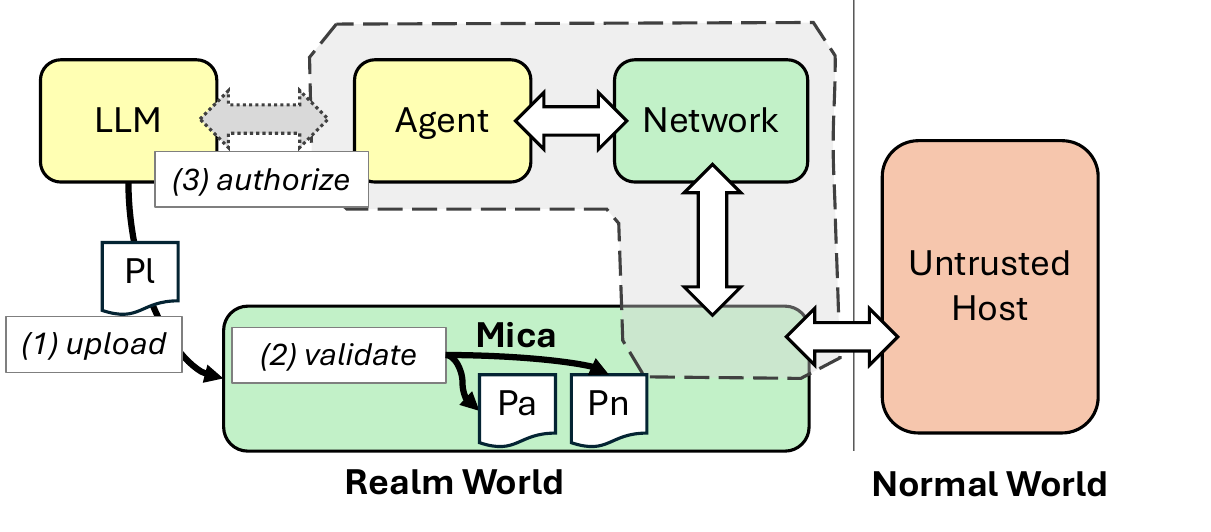}
  \caption{\system{} validates the LLM's policy ($P_L$) against peer policies ($P_A$, $P_N$) before group admission (grey). White arrows denote active authorized channels; grey arrows denote channels pending validation. Green components may communicate outside the group.
  }
  \label{fig:mica-overview}
\end{figure}

%% file: sections/rmm_impl.tex

\label{sec:rmm_impl}

We implement \system{} on top of Arm Confidential Compute Architecture (CCA).
Arm CCA provides an extensible substrate with the primitives required by \system{}, enabling a firmware-based implementation without hardware modifications.
Appendix~\ref{sec:appendix:cca} provides a detailed description of CCA's main components.
Our implementation targets the QEMU CCA platform~\cite{qemu-repo} and the OpenCCA~\cite{opencca} platform, as CCA-capable silicon is not yet widely available.


\system{} is implemented through changes to:
(1) the Realm Management Monitor (RMM) (TF-RMM~\cite{rmm-repo}) to support policy specification and enforcement,
(2) QEMU (v9.2.4~\cite{qemu-repo}) VMM to expose these mechanisms, and
(3) the CCA-enabled guest and host kernels~\cite{linux-repo} to leverage \system{}’s API.
\autoref{tbl:ext-loc} includes the lines of code added to each component.


This section focuses on the TF-RMM extensions, while \autoref{sec:infrastructure} describes the supporting infrastructure in KVM and QEMU.

\input{tables/cloc-slim}
\input{tables/new_apis-slim}

\subsection{Overview}

\system{} integrates closely with the standard CCA control flow to minimize disruption.
In CCA, most management operations are delegated to the untrusted host.
\system{} preserves this model, while extending it with calls from the host to the RMM (RMI calls) for resource management and channel setup, and from Realm to the RMM (RSI calls) for policy upload and attestation.
\autoref{tbl:new-api} summarises the added RMIs and RSIs.
We present the implementation in the order of steps in \systems Realm execution workflow.


\subsection{Creation and Pre-population}
\label{sec:impl:create}

\system{} introduces two minor extensions to the standard Realm creation flow:
(1) delegate a granule to store Realm policy metadata (PD), and
(2) delegate and pre-mark protected granules that will back communication channels.
These changes ensure that a Realm intending to use \system{} reaches a well-defined state from which it can upload a policy and subsequently participate in a group.

\myparagraph{Policy metadata}
\system{} extends the RMM to maintain a policy configuration for each Realm.
To support this, the RMM requires exclusive access to protected memory to store the policy metadata.
\system{} reuses the existing \texttt{RMI\_GRANULE\_DELEGATE}, which allows the untrusted host to delegate ownership of a granule to the Realm world.
The host invokes the new \texttt{RMI\_REALM\_PD} to designate the delegated granule as policy metadata (PD) for a specific Realm.
Initially, the PD encodes an empty policy configuration.
The RMM records the PD granule in the RD using a new 64-bit field, \texttt{pd}, which stores its PA.
The PD is managed exclusively by the RMM and is never mapped into the Realm’s stage-2 address space.

\myparagraph{Allocating and pre-marking protected channel granules}
\system{} extends the CCA protocol for protected memory to explicitly identify granules that may be shared across Realms, \ie granules intended to back protected channels.
On the first Realm's memory access fault caused by a missing stage-2 mapping, the RMM exits to the host, which allocates backing physical memory and delegates the corresponding granule to the Realm world via \texttt{RMI\_GRANULE\_DELEGATE}, before establising the stage-2 mapping and re-entering the Realm.
\system{} then extends this flow with a new RMI.
The host invokes \texttt{RMI\_DATA\_CREATE\_UNKNOWN\_SHARED} to designate the granule as shared and map it into the Realm.
This call has the following effects:
(1) if this is the first time the granule is used for sharing, it marks the granule as \emph{shared} and initializes \systems per-granule metadata;
(2) it records the Realm identifier in this metadata;
(3) it installs the IPA-to-PA mapping in the Realm’s second-stage translation tables (RTTs). 
At this point, the RMM installs the RTT entry with no access permissions, so any access to the IPA triggers a permission fault.
\system{} enables the appropriate access rights only after the Realm uploads its policies.

\system maintains a table that includes all the shared granules, the \textit{Shared Granule Table} (SGT).
For each granule, the SGT keeps the Realm identifiers that map the granule, along with the associated access rights.
The RMM uses this table for policy validation and during the attestation to find the accessible peers via the active channels.
Following CCA’s typed-memory model, \system{} introduces a new RMI (\texttt{RMI\_SGT}) to designate granules for hosting the SGT.

For unprotected channels, \ie shared with the host, the allocation and mapping of memory follows the standard CCA protocol for unprotected shared memory.

\subsection{Supplying Realm Policies}
\label{sec:impl:policy}

After boot, a Realm needs to upload its policy configuration before it can join a \emph{group} and start interacting with peers.

\myparagraph{Policy upload}
A Realm uploads its policy using \system{}'s \texttt{RSI\_UPLOAD\_POLICY}, specifying a protected IPA pointing to a binary-encoded policy configuration. 
Upon receiving this RSI, the RMM copies the binary blob into the PD.

\myparagraph{Validation}
The RMM validates the policy blob to ensure that all memory accesses comply with the configuration. 
First, the RMM unmaps all the unprotected channels which are not explicitly covered in the policy. 
For protected channels, the RMM performs the validation algorithm described in \autoref{sec:design:pve}: it first checks that, for each protected channel in the policy, all IPAs are backed by shared granules recorded as SGT entries, and then performs a graph traversal over all channels that would become active -- i.e., channels where a peer exists and has supplied its policy.
If any violation is detected, the operation fails.
In the current implementation, the Realm is terminated for simplicity, though an alternative would be to reject the policy and signal an error to the Realm.


If validation succeeds, the RMM updates the corresponding REM~\cite{cca} with a digest of the validated policy, making it part of the attestation report. 
Since REM fields already support deployment-specific extensions, \system{} required minimal changes to integrate policy measurement into attestation.

Finally, the RMM modifies the access rights to mappings that correspond to protected channels and thus enables Realm access to these memory regions.
Furthermore, from this point onward, the RMM will consult the PD whenever a control-flow channel is trigger by the Realm's actions and decide whether to allow, scrub, or block this transition to the host.
Note that once a Realm's policies have been validated, \system{} freezes it and does not require re-validation later on.
Moreover, \system ensures that the policies supplied subsequently by other peers do not contradict with the currently validated ones.
This includes checking consistency across peers, such as matching \emph{type} and \emph{size}, compatible access rights, and non-conflicting transition policies.

\subsection{Attestation}
\label{sec:impl:runtime}

To support group attestation, \system{} introduces two RSIs (\texttt{RSI\_ATTEST\_TOKEN\_INIT\_GROUP} and
\texttt{RSI\_ATTEST\_TOKEN\_CONTINUE\_GROUP}), adapting existing attestation semantics in \arch{} CCA~\cite{cca}.
Group attestation retains the verifier-provided nonce and RMM measurement, allowing verifiers to reject replayed reports and detect RMM modifications; snapshot restoration, Realm termination, and reboot invalidate the attested group and require forming and attesting a new one.
Our prototype scans the Shared Granule Table for every attestation request, instead of caching validated groups' structure.
We favor this trade-off because memory is limited in our QEMU setup and attestation is infrequent, occurring only before secret provisioning or when establishing trust with the Realm.
Caching the validated group structure in RMM memory would avoid rebuilding it for each attestation, reducing processing time, but it would require additional RMM memory.

%% file: tables/cloc-slim.tex
\begin{table}[t]
\centering
\scriptsize
\setlength{\tabcolsep}{2pt}
\renewcommand{\arraystretch}{0.92}
\begin{tabular}{|l|c|}
\hline
\textbf{Component} & \textbf{Added LoC} \\ \hline
RMM (cca/v8)~\cite{rmm-repo} & 3723 (7\%) \\ \hline
Host kernel~\cite{linux-repo} & 537 (<1\%) \\ \hline
QEMU VMM (cca/2025-06-12)~\cite{qemu-repo} & 219 (<1\%) \\ \hline
Guest kernel~\cite{linux-repo} & 1786 (<1\%) \\ \hline
\end{tabular}
\caption{Lines of code added to each component in \system.}
\label{tbl:ext-loc}
\end{table}

%% file: tables/new_apis-slim.tex
\begin{table}[t]
\centering
\scriptsize
\setlength{\tabcolsep}{2pt}
\renewcommand{\arraystretch}{0.92}
\resizebox{\columnwidth}{!}{%
\begin{tabular}{|l|l|}
\hline
\multicolumn{2}{|c|}{\textbf{RMIs}} \\ \hline
\texttt{RMI\_DATA\_CREATE\_UNKNOWN\_SHARED} & Mark protected-channel PAs shared across realms \\ \hline
\texttt{RMI\_REALM\_PD} & Mark page as PD \\ \hline
\texttt{RMI\_SGT} & Mark page as SGT \\ \hline
\multicolumn{2}{|c|}{\textbf{RSIs}} \\ \hline
\texttt{RSI\_UPLOAD\_POLICY} & Upload realm policy to RMM \\ \hline
\texttt{RSI\_ATTEST\_TOKEN\_INIT\_GROUP} & Start group attestation \\ \hline
\texttt{RSI\_ATTEST\_TOKEN\_CONTINUE\_GROUP} & Continue group attestation retrieval \\ \hline
\end{tabular}%
}
\caption{New APIs added in \system.}
\label{tbl:new-api}
\end{table}

%% file: sections/infrastructure.tex
\label{sec:infrastructure}

\system{} requires minor extensions to the software stack responsible for creating and managing Realms.
We modified the KVM~\cite{kvm}/QEMU~\cite{qemu} VMM stack for the untrusted host and added a simple policy driver to Realm guest Linux kernels.
These changes are small, integrate cleanly with the existing infrastructure, and preserve the standard \arch{} CCA control flow while enabling attested channel support.

\myparagraph{KVM Extensions}
Recent versions of KVM~\cite{kvm} include support for CCA, allowing an untrusted Linux host to create and manage Realms.
In particular, KVM exposes an API to mark a Realm’s memory as protected by associating the \texttt{KVM\_MEMORY\_ATTRIBUTE\_PRIVATE} flag with a userspace memory slot assigned to the Realm.
Conversely, memory slots without this attribute are treated as unprotected. 

\system{} extends this mechanism with a new memory attribute, \texttt{KVM\_MEMORY\_ATTRIBUTE\_PRIVATE\_SHARED}, which identifies protected memory regions intended to back protected channels.
KVM treats these memory regions as shared between realms. It delegates the corresponding granules to the Realm world with \texttt{RMI\_GRANULE\_DELEGATE} and makes them accessible to each realm participating
realm with  \texttt{RMI\_DATA\_CREATE\_UNKNOWN\_SHARED}.




Finally, we modify KVM to allocate a page, by default, for every Realm to serve as the PD granule and one page for the granule that will hold the SGT, and perform the corresponding RMIs to coordinate with the RMM.
In the current implementation, a single page is enough to host Realm policies but should the Realm require more metadata, the RMM can signal the host to provide more granules via error code as return values to RMIs.


Overall, the required KVM changes are modest and localized, serving primarily to surface \system{}’s semantics to the host while leaving enforcement to the RMM.

\myparagraph{Qemu VMM Extensions}
QEMU~\cite{qemu} provides extensive device emulation and supports emerging hardware technologies, including recent extensions for \arch{} CCA~\cite{qemu_support_for_cca}, enabling the execution of Realms on compatible \arch{} platforms.
It also, exposes Inter-VM Shared Memory (\emph{ivshmem})~\cite{ivshmem} to enable direct communication via unprotected shared memory between multiple VMs or between a VM and the host.
Ivshmem is backed by unprotected shared memory and accessed by QEMU via the host file system.
VMs requiring shared access are instantiated with ivshmem devices pointing to the same backing file, instructing QEMU to map the corresponding memory region into each VM.
Within the guest, ivshmem appears as an emulated PCI device with an associated (guest) physical memory region for MMIO.
Guest operating systems provide drivers to manage the device and expose its memory to applications through the file system.

To seamlessly support \system{}'s protected and unprotected channels, we extended QEMU’s ivshmem implementation to accept an additional \emph{protected} flag.
When set, this flag instructs QEMU to report the corresponding memory region to KVM using the \texttt{KVM\_MEMORY\_ATTRIBUTE\allowbreak\_PRIVATE\_SHARED} attribute, marking it as backing a protected channel.
This allows us to reuse QEMU's existing ivshmem infrastructure to support protected channels.
KVM will then proceed with the workflow described above, delegating the granule, marking it as protected shared, and mapping it into the Realms address spaces using \system{} RMIs.

\myparagraph{Guest Realm Extensions}
On the guest side, we extend the Realm guest kernel to support policy upload, \systems{} group attestation, and all the interactions with the RMM.
To ease policy upload, we implement a guest kernel module that exposes a character device at \texttt{/dev/rsi\_upload}.
The driver receives the write requests and accumulates the policy bytes into an internal buffer.
The driver pre-validates the policy and can optionally pre-fault the Realm's address space.
It then translates the JSON into the binary format expected by the RMM, and issues an \texttt{RSI\allowbreak\_UPLOAD\_POLICY}, passing the guest physical address of the binary blob.
We also, extend the guest-kernel attestation driver to support \system{} group attestation.
It receives attestation requests via an \texttt{ioctl} call and in turn performs the appropriate \texttt{RSI\_ATTEST\_TOKEN\allowbreak\_INIT\_GROUP} and \texttt{RSI\_ATTEST\allowbreak\_TOKEN\allowbreak\_CONTINUE\_GROUP} (\autoref{sec:impl:runtime}).

\myparagraph{Alloy Model Checking}
We formalize the semantics of \system's configuration language using Alloy~\cite{jackson2002alloy}, which lets us describe sets of peers and channels and constrain their relationships.
This formalization serves two purposes.
First, it formalizes the semantics of complex configuration options and clarifies their implications, \eg what it means for a peer to be \texttt{strict} or a channel to permit \texttt{ANY} peers.
Second, the Alloy Analyzer lets us check properties of concrete configurations, \eg as whether a potential communication path exists between a peer and the host or whether specific peer configurations are mutually compatible.
To support these checks, we implemented a translator that maps the JSON policies used by peers in \system to Alloy predicates.
Alloy's relational model fits \system naturally: we encode peers and channels as \texttt{Signatures}, which define sets of elements, and describe their allowed relationships using \texttt{Facts}, which express constraints that every valid model instance must satisfy.

%% file: sections/evaluation_asplos.tex
\label{sec:eval}


This section evaluates \system{} along two dimensions: the overhead of its extensions to Arm CCA and its expressiveness in supporting realistic confidential workloads.

We first quantify the potential runtime overheads and performance gains introduced by \system{} by benchmarking its implementation atop OpenCCA \cite{opencca}, an open-source performance emulator for Arm CCA.
Specifically, we measure the overhead added by \system{} in Realm initialization and attestation, and the improvements in group communication comparing against vanilla CCA.

We then use representative application scenarios implemented in our QEMU-based functional prototype to qualitatively demonstrate \system{}, where complex multi-Realm pipelines can be exercised and pushed further than on current OpenCCA platforms.
These scenarios illustrate the expressiveness of \systems policies in capturing real-world communication patterns across mutually distrustful components, and enable analysis of their impact on the trusted computing base and resulting security benefits.



\subsection{Performance Prototype}

\input{images/eval/attestation_opencca}
\input{images/eval/message_passing_opencca}

We evaluate the overhead of \system{} by benchmarking its implementation using the OpenCCA~\cite{opencca} framework, which approximates Arm CCA performance in the absence of widely available hardware, and analyze setup, communication, and attestation costs.
All performance experiments are conducted on a Radxa Rock 5B~\cite{radxa-rock5bp} with 16GB of RAM and an 8-core processor (4× ARM Cortex-A76 and 4× ARM Cortex-A55), running each experiment 10 times and reporting averaged results.
Our evaluation uses Realms with 512\,MB memory and 256\,KB (64-page) protected channels, and is designed to test three metrics:
(i) the setup overhead introduced by \system{},
(ii) the overhead of \systems group attestation in terms of latency and size,
(iii) and the runtime benefits of \systems protected channels over encrypted host-mediated communication.

\myparagraph{Setup Overhead}
\system{} introduces modest setup overhead due to additional RMIs, policy management, and protected channel initialization.
We observe an average increase of 2--4\% in Realm boot time due to additional \system RMIs, along with approximately 1.5\,s to upload and verify policies.
Establishing protected shared memory incurs higher cost than host-backed mappings, as our implementation requires delegating fresh granules to the Realm world: setting up a protected \texttt{ivshmem} channel takes $\sim$0.06\,s per page on average, compared to $\sim$0.012\,s for non-protected \texttt{ivshmem}.
While this increases setup latency, it is a one-time cost and enables secure communication between Realms via protected memory inaccessible by the host.

\myparagraph{Attestation Overhead}
\system{} incurs moderate overhead in group attestation latency and report size as the number of Realms increases, while exposing richer information about the structure of the confidential pipeline than standard single-Realm CCA attestation (Fig.~\ref{fig:att-microbenchmark}).
Compared to single-Realm attestation, two-Realm group attestation increases latency by approximately 34\%, and three-Realm attestation by around $3\times$, due to the prototype's quadratic iteration over SGT entries during group attestation; future work will address this with more efficient traversal and indexing strategies.
Similarly, report sizes grow by roughly 35\% and 60\% for two- and three-Realm groups, respectively. 
Despite these increases, attestation remains practical and scales reasonably with group size, enabling end-to-end verification of multi-Realm deployments. 
In contrast, equivalent coverage in vanilla CCA would require a client to attest each component individually, without evidence of how those components interact.


\myparagraph{Communication Overhead}
Once a channel is activated, \system{} uses protected shared memory, similar to CAEC's data path, without further policy checks or RMM processing.
We implement message passing directly over a 256\,KB shared memory region and measure one-way sender-to-receiver latency per message.
Messages are split into frames with metadata (e.g., message id, chunk index, size, sequence number), written to shared memory, and then polled, reconstructed, and ACKed by the receiver.
As in CAEC, \system{} transmits data in plaintext within protected memory. For context, we compare this data path with communication over unprotected memory using two OpenSSL encryption modes: a lightweight mode (AES-256-CTR + HMAC-SHA256) and a heavier mode
(AES-256-CBC with PBKDF2 and HMAC-SHA256).
Across all configurations and message sizes, \system{} strengthens CAEC's security guarantees without additional per-message processing, preserving its $2.5\times$--$14.5\times$ latency advantage over these encrypted modes (Fig.~\ref{fig:msg-microbenchmark}).

\subsection{Functional Prototype}

\input{images/usecases_slim}


This subsection uses an emulation-based functional prototype to demonstrate \systems expressiveness and analyze its impact on the TCB and security properties.

We evaluate \system{} through four representative and increasingly complex use cases, each illustrating a distinct communication and trust pattern across local and cloud confidential workloads:
(1) a network-mediated local agent pipeline, where all external access is routed through a dedicated trusted networking service,
(2) a local confidential inference pipeline, where the agent invokes a reusable LLM service while external communication remains mediated by the trusted networking service (similar to the motivating example in \S\ref{sec:motivation}),
(3) a guardrailed local confidential inference pipeline, where the agent composes LLM and guardrail services while all external communication remains mediated by the trusted networking service,
and (4) a confidential video moderation pipeline of mutually distrustful components; \autoref{fig:usecases} summarizes their structure and interactions.
These scenarios are implemented in QEMU 9.2.4~\cite{qemu-repo}, which enables flexible multi-Realm experimentation, although its current 3GB host memory limit constrains the number and size of concurrent Realms.
In some cases, this necessitates the use of virtual disks instead of memory-resident ramdisks, an artifact of the emulation environment that would not arise on real hardware; where relevant, we describe the intended deployment without these constraints.




\myparagraph{Network-Mediated Agent (Fig. \ref{fig:usecases}a)}
This use case captures a local agent pattern in which a confidential agent with dynamic, evolving control logic accesses external networked resources only through a small trusted gateway.
This separates sensitive agent logic from the host-facing communication boundary and gives the agent owner tighter control over external access: the agent may evolve to invoke new tools or compose different requests, while the gateway continues to enforce fixed destination filtering and policy checks.
We implement this pattern as a two-Realm pipeline \realm{Agent} $\leftrightarrow$ \realm{Net}, where \realm{Agent} runs a local agent implemented with \texttt{LangChain}~\cite{langchain} and \realm{Net} acts as a trusted networking gateway implemented with \texttt{raw sockets}.
\realm{Net} exposes a reusable networking interface using the \systems \texttt{any} identifier, allowing authorized agents to bind to the service without depending on a specific gateway instance.
\realm{Agent} sends external-access requests over a protected memory channel, while \realm{Net} validates destinations, mediates communication with external resources, and releases data only to policy-authorized endpoints.
\system{} enforces the mediated structure by construction: the agent may communicate with \realm{Net}, but may not directly map host-network resources or unintended interfaces.
Thus, host-facing communication is centralized in a small trusted component, while confidential agent logic remains isolated from direct host interaction.

\myparagraph{Network-Mediated Agent with Local LLM (Fig. \ref{fig:usecases}b)}
This use case adds local confidential inference to the network-mediated agent pattern, allowing the agent to invoke a local LLM while external communication remains controlled by the trusted networking service.
This is useful when the model represents intellectual property or when inference stability matters: agents may use the model without exposing its implementation or weights to agent owners, or depending on an LLM API whose served model may change over time~\cite{llmapistability}.
We implement this by adding a model Realm \realm{Model} to the previous pattern, forming \realm{Model} $\leftrightarrow$ \realm{Agent} $\leftrightarrow$ \realm{Net}.
\realm{Model} runs local inference using \texttt{llama.cpp}~\cite{llamacpp}; \realm{Agent} invokes it over an explicitly authorized protected memory channel, while external-access requests continue through \realm{Net}.
\system{} enforces this structure by allowing \realm{Agent} to communicate only with the authorized model and networking interfaces.
As a result, model execution remains isolated from host-facing communication, while the agent composes local inference with controlled external access.

\myparagraph{Network-Mediated Agent with Guardrails (Fig. \ref{fig:usecases}c)}
This use case adds a local confidential guardrail service to the agentic inference pattern, ensuring that prompts and model outputs are mediated before any result is released externally.
Such structures arise when the agent relies on a local LLM, but policy, safety, or formatting constraints must be enforced by a separate component rather than by the agent or model itself.
We implement this by adding a guardrail Realm \realm{Guard} between \realm{Agent} and \realm{Model}, forming \realm{Model} $\leftrightarrow$ \realm{Guard} $\leftrightarrow$ \realm{Agent} $\leftrightarrow$ \realm{Net}.
\realm{Guard} mediates requests and responses over explicitly authorized protected memory channels, while \realm{Net} remains the only Realm handling external communication.
\system{} enforces this structure by allowing \realm{Agent} to communicate only with the authorized guardrail and networking interfaces, and \realm{Model} only with \realm{Guard}.
As a result, neither raw prompts nor raw model outputs can bypass mediation, while all external communication remains confined to the trusted networking service.

\myparagraph{Multi-Stage Video Moderation (Fig. \ref{fig:usecases}d)}
This use case captures a common cloud pattern of strictly feed-forward dataflow across mutually distrustful components, where each stage performs a transformation without learning others’ internals and no backward or lateral communication is permitted.
Such pipelines are widely used in media processing, analytics, and compliance workloads.
We implement this pattern as a three-Realm pipeline \realm{Net} $\rightarrow$ \realm{Enc} $\rightarrow$ \realm{Nudity} $\rightarrow$ \realm{Net}, where a gateway \realm{Net} handles I/O, an encoder \realm{Enc} processes video using \texttt{ffmpeg}~\cite{ffmpeg}, and a moderation stage \realm{Nudity} analyzes content using \texttt{nudenet}~\cite{nudenet}.
\realm{Net} ingests encrypted input, distributes decrypted data through protected channels, and collects the final output for encryption and storage. All components run unmodified inside CVMs.
\system{} enforces the feed-forward structure by construction: each stage may read only from its predecessor and write only to its successor, while all other mappings are disallowed.
Intermediary Realms have no access to the host or external interfaces, preventing data exfiltration.
Thus, end-to-end confidentiality is enforced by policy, independent of component correctness or trustworthiness.

\myparagraph{TCB and Security Analysis}
\system{} reduces the trusted computing base (TCB) by confining data exfiltration capabilities to a small set of policy-authorized components.
In our deployments, only gateway Realms that interface with the untrusted host or external network are part of the TCB, while application Realms remain outside despite processing sensitive data, as policies prevent them from communicating beyond explicitly allowed channels.
This distinction is particularly important for agentic workloads: agent code is complex, dynamic, and difficult to meaningfully measure, but its network effects can still be controlled by forcing all external communication through a small measured networking Realm.
By contrast, constraining the agent inside an in-TEE sandbox~\cite{ryoan} or paravisor-based runtime~\cite{veil,erebor,verismo} would require adapting the agent to the constrained environment and may expose the restriction boundary to the agent itself, increasing the risk that it observes, reasons about, or acts adversarially against the constraints~\cite{aisi}. As shown in Table~\ref{tbl:ext-loc}, \system increases the RMM TCB by only 7\%, yet reduces the application TCB by roughly $2$–$3\times$ compared to vanilla CCA, where all communicating TEEs must be trusted.

Beyond TCB reduction, \system{} enforces communication topology as a policy property, rather than relying on components to voluntarily follow a protocol.
This yields two kinds of security guarantees in our use cases.
First, in the video moderation pipeline, \system{} enforces feed-forward information flow: a compromised encoder can access plaintext inputs but cannot exfiltrate them, as policies permit it to communicate only with the next stage and disallow host or external interfaces.
Second, in the guardrailed LLM agentic pipeline, \system{} enforces mandatory mediation: external communication must pass through \realm{Net}, and in the guardrailed case \realm{Model} communicates with \realm{Agent} only through \realm{Guard}.
Encrypted channels alone can protect channel contents~\cite{ryoan}, but they do not impose this topology: they remain compatible with direct bidirectional communication, bypass paths, or unintended host access.
By contrast, \system{} enforces feed-forward flow, mandatory mediation, and absence of host access at the platform level through policy.

%% file: images/eval/attestation_opencca.tex
\begin{figure}
    \centering
        \includegraphics[width=0.8\linewidth]{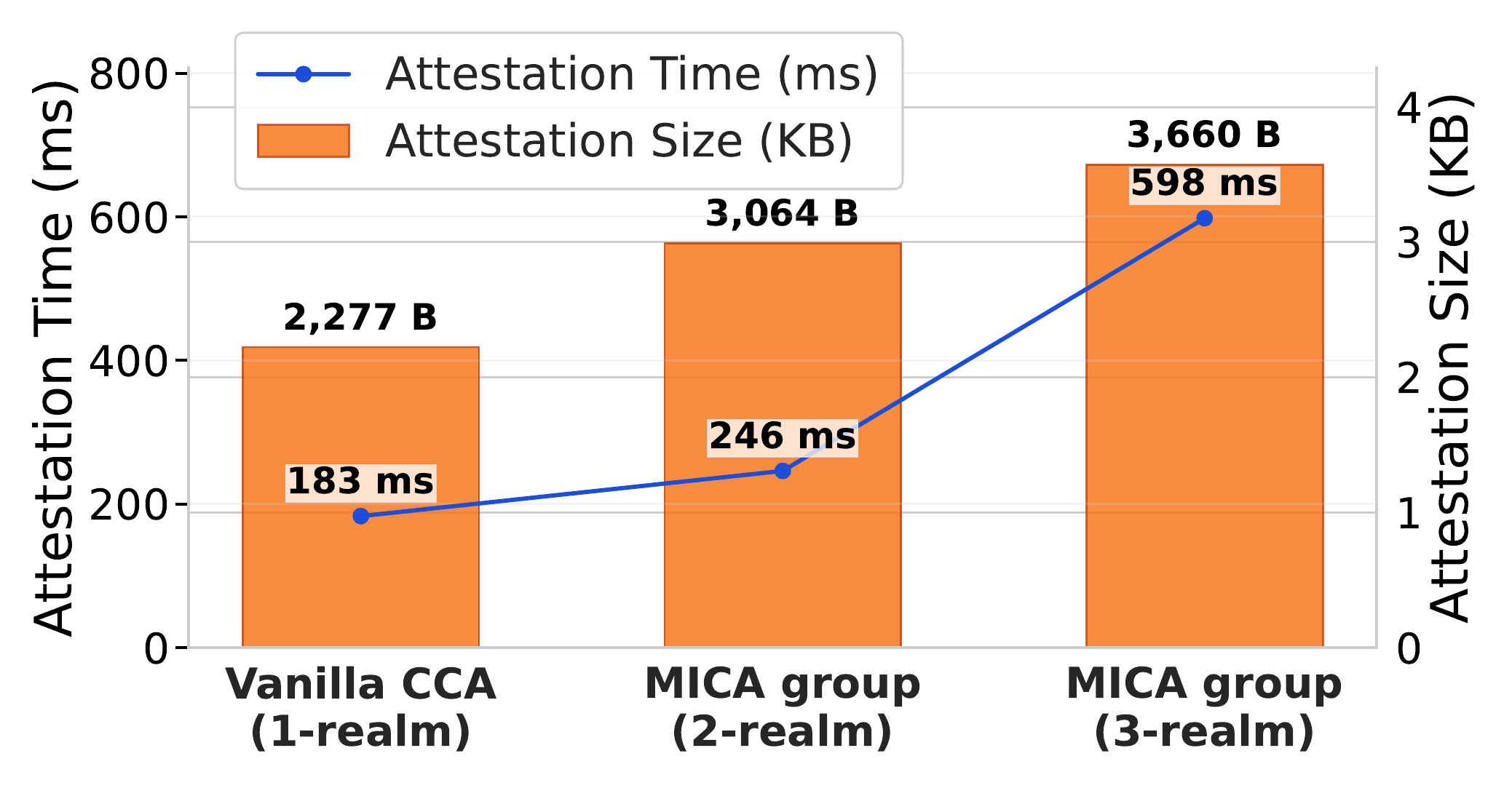}
        \caption{Attestation report size (KB, barplot) and retrieval latency (ms, lineplot), comparing vanilla CCA (1 realm) and \system group attestation (2–3 realms).}
        \label{fig:att-microbenchmark} 
\vspace{-10pt}
\end{figure}


%% file: images/eval/message_passing_opencca.tex
\begin{figure}
    \centering
        \includegraphics[width=0.8\linewidth]{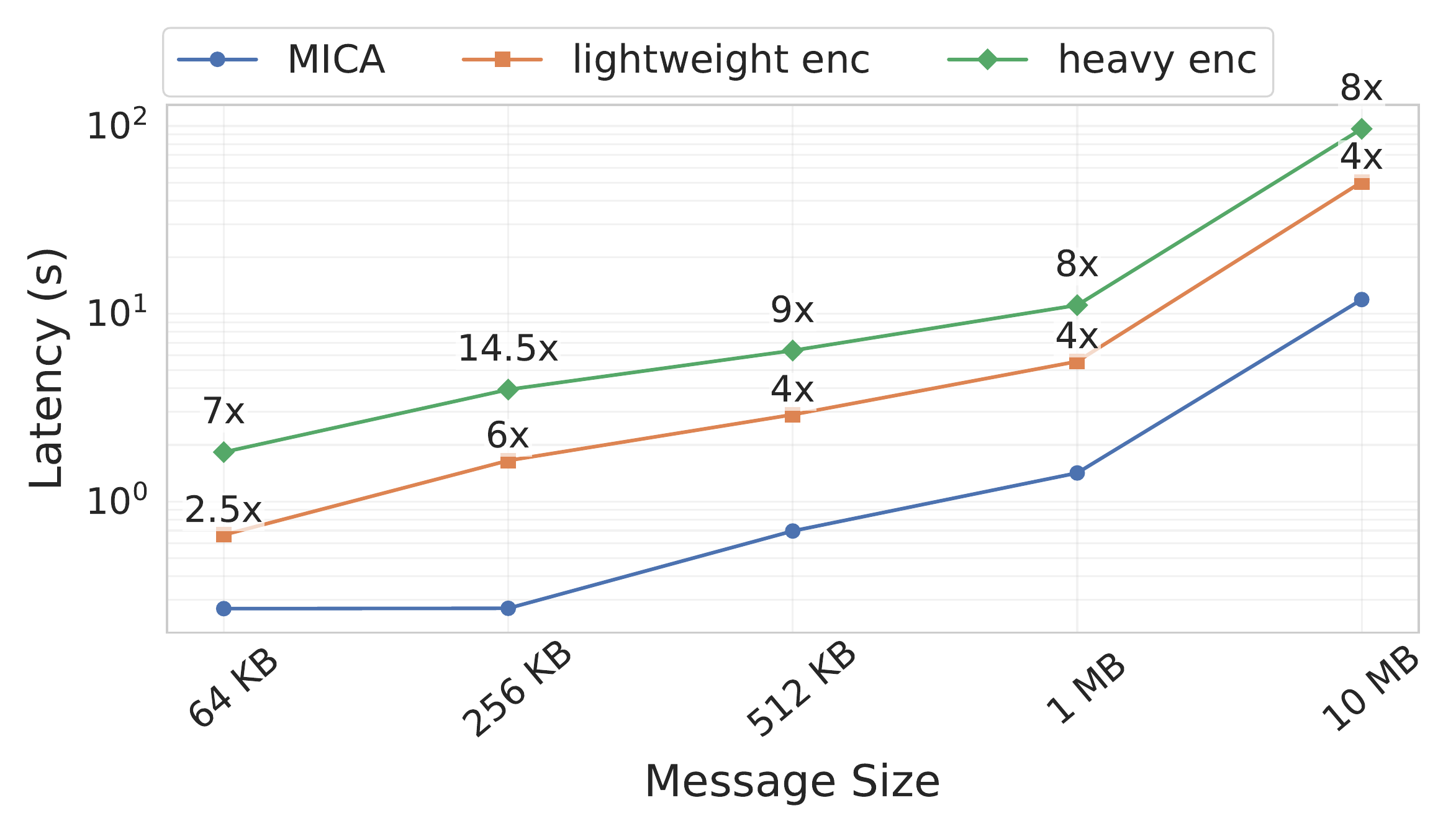}
        \caption{Message-passing latency (s) over shared memory across payload sizes. \system{}'s protected-memory data path achieves a $2.5\times$--$14.5\times$ latency advantage over encrypted host-mediated communication.}
        \label{fig:msg-microbenchmark} 
\vspace{-10pt}
\end{figure}


%% file: images/usecases_slim.tex
\begin{figure*}
  \centering
  \includegraphics[width=0.8\textwidth]{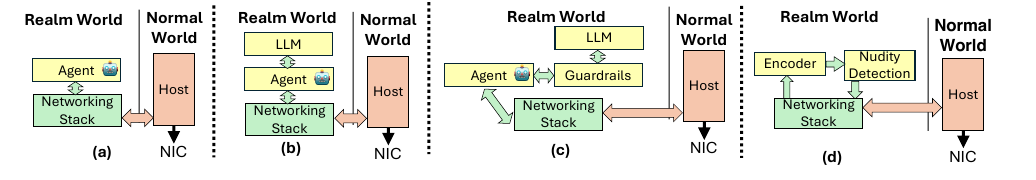}
  \caption{
  Application scenarios over \system. Green denotes trusted/confidential components, red untrusted components, and yellow confidential but distrustful components. Incoming($\rightarrow$)/outgoing($\leftarrow$) arrows denote read/write access.}
  \label{fig:usecases}
  \vspace{-10pt}
\end{figure*}

%% file: sections/related_work_asplos.tex


\myparagraph{Sharing memory between TEEs}
Prior work explored mechanisms for protected memory sharing between TEEs. 
Plug-in Enclave~\cite{plug-in-enclave} enables read-only shared enclaves on Intel SGX~\cite{sgx}, Cerberus~\cite{cerberus} verifies immutable sharing on RISC-V Keystone~\cite{keystone}, and Elasticlave~\cite{elasticlave} supports writable shared memory on RISC-V. 
CHERI~\cite{cheri} similarly allows compartments to share a single address space. 
CAEC~\cite{sinas-arxiv} implements shared memory in \arch{} CCA, with sharing initiated and driven by the Realms. 
\system{} extends and improves this design by decoupling memory mapping from memory activation: shared regions can be established first, then activated only after RMM validation of user-specified policy. 
This enables policy-driven sharing and attestation.



\myparagraph{Constraining communication}
Ryoan~\cite{ryoan} confines confidential data to NaCl~\cite{nacl} and Intel SGX~\cite{sgx} sandboxes, preventing exfiltration by untrusted processing components.
Confidentiality relies on a trusted runtime that controls data access and flow.
Other works use double-sided sandboxing to mitigate leakage, including side channels~\cite{hse,iovalve}.
Veil~\cite{veil} introduces enclave-like isolation inside AMD SEV-based CVMs, while Erebor~\cite{erebor} provides sandboxing for Intel TDX~\cite{tdx}.
Software-based approaches~\cite{inktag, overshadow, hyperenclave} can also restrict sensitive data access to a small trusted software base within a CVM.
\system{} instead enforces constraints \emph{from outside the TEE} without modifying application code.
Tyche~\cite{tyche, tyche2} explores security-monitor-based enforcement of strong isolation and \emph{controlled sharing} between mutually distrustful components.


\myparagraph{Attesting multiple compartments}
Recent works on swarm and collective attestation~\cite{prive, prove, zekra} explore mechanisms for attesting groups of TEEs.
PRIVÉ~\cite{prive} enables anonymous attestation, PROVE~\cite{prove} supports public verification by uncoordinated verifiers, and ZEKRA~\cite{zekra} enforces control-flow integrity via zero-knowledge proofs.
Unlike \system{}, these approaches assume TEEs are trusted once attested.



\myparagraph{Microkernels}
Microkernels such as seL4~\cite{sel4} and HongMeng~\cite{hongmeng} isolate services and mediate communication to limit compromised components' authority. \system applies this principle to application logic, partitioning it into mutually distrusting components with enforced communication policies.

\myparagraph{MAC and IFC}
MAC systems such as SELinux~\cite{selinux, selinux-paper}, AppArmor~\cite{apparmor, apparmor-paper}, and Flask~\cite{flask} control access to OS resources, while \system enforces coarser policies at the application-component level. IFC systems such as HiStar~\cite{histar} and Asbestos~\cite{asbestos} track individual data flows; \system instead attests and controls communication channels without tracking data propagation.

%% file: sections/conclusion.tex
\system{} decouples confidentiality from trust in in-TEE code.
By making shared channels between TEEs and with the untrusted host explicit, constrained, and attested, \system{} alleviates the need to trust the functional correctness of individual components processing sensitive data.
Its transitive policies and attestation cover entire computation pipelines made of distrustful components.
Our \arch{} CCA prototype shows that this model supports modern confidential agentic applications and pipeline computations with modest overheads.

%% file: sections/acknowledgments_final.tex
We thank the anonymous reviewers and our shepherd for their valuable feedback. 
This research was supported by the AI Security Institute (AISI) Systemic Safety Grants Programme, UKRI Open Plus Fellowship (EP/W005271/1 Securing the Next Billion Consumer Devices on the Edge), and an Amazon Research Award ``Auditable Model Privacy using TEEs''.


%% file: sections/policy_appendix.tex
\section{Example of policy file}
\label{sec:appendix:policy}
The following example illustrates a policy file defining peers and a shared memory region. It highlights how components specify peer identities and shared channels.
\input{images/policy}

%% file: images/policy.tex
\begin{figure}[h]
    \centering
    \begin{lstlisting}[language=json,basicstyle=\ttfamily\footnotesize]
{
  "Peers": {
    "Self": "P1",
    "P1": { "hash": 0xabcd, "is_gateway": true, "strict": false},
    "P2": { "hash": 0xefab, "is_gateway": false, "strict": true},
  },
  "MemChannels": {
    "Mem1": {
      "size": 0x4000000,
      "type": "PROTECTED",
      "mappings": {
        "P1": { "gpa": 0x18000000000, "prot": "W" },
        "P2": { "gpa": 0x18000000000, "prot": "R" },
        "ANY": { "gpa": 0x18000000000, "prot": "RW", "count": -1 },
      }
    },
  },
  "TransChannels": {
    "CF1": {
      "owner": "P2",
      "type": "exception"
      "range": ["1", "4"],
      "policy": "ALLOW"
    },
    "CF2": {
      "owner": "P1",
      "type": "call"
      "range": ["0"],
      "policy": "SCRUB"
    },
  }
}
\end{lstlisting}
\caption{Example of policy file used by \system defining two peers with one shared memory region. P1 is the component supplying the policy as indicated by \texttt{Self}.}
    \label{fig:policy}
\end{figure}


%% file: sections/cca_bg_appendix.tex
\section{Background on \arch{} CCA}
\label{sec:appendix:cca}

\arch{} Confidential Compute Architecture (\arch{} CCA) is a set of Armv9-A architectural extensions that provide confidential virtual machines (CVMs).
CCA introduces two new execution worlds, the Root world and Realm world, alongside the existing Normal world and Secure world used by TrustZone~\cite{trustzone}.
worlds define isolation domains that coexist with the architectural privilege levels (EL0–EL2): in all worlds, applications run at EL0, operating systems at EL1, and hypervisors/firmware at EL2.
The highest privilege level, EL3, hosts the Monitor, which executes in the Root world.
The Monitor is responsible for switching between worlds and managing world-level state.
\autoref{fig:cca-arch} shows the overall architecture of \arch{} CCA.

CCA enforces strict memory ownership at physical-page granularity (i.e., per granule) by associating each granule with a world state that reflects which world owns it at a given time.
Granules tagged as Normal are accessible from all worlds, whereas granules tagged as Secure or Realm are accessible only from their respective worlds and from the Root world.
These tags are stored in the Granule Protection Table (GPT), a hardware structure managed exclusively by the Monitor at EL3.
During address translation, the hardware performs a Granule Protection Check (GPC) on the GPT to ensure the access is permitted for the currently executing world.

\input{images/cca-arch}

The Realm world hosts CVMs, known as \emph{Realms}, whose memory and execution state are protected from the host OS and hypervisor in Normal world, and from software running in Secure world.
Realms run in EL1 and EL0, while a Realm Manager Monitor (RMM) acts as a trusted firmware in Realm world's EL2.
The RMM controls the life cycle of Realms.
It handles memory management via the \textit{Realm Translation Tables} (RTTs), which map \textit{Intermediate Physical Addresses} (IPAs) to \textit{Physical Addresses} (PAs) -- effectively performing guest-physical to host-physical address translation, and mediates all interactions with the untrusted hypervisor in Normal world.
The RMM maintains per-Realm metadata in a dedicated datastructure called the \textit{Realm Descriptor} (RD).

Unlike CISC architectures such as AMD SEV-SNP, which rely on a closed-source Platform Security Processor (PSP) and restrict interaction to a fixed set of domain-specific instructions, CCA follows the RISC philosophy: it builds on a minimal set of micro-architectural components and supports an extensible RMM design.
The RMM mediates all interactions between the Realm and Normal worlds through two interfaces: the \textit{Realm Management Interface} (RMI) and the \textit{Realm Services Interface} (RSI).
The hypervisor communicates with the RMM to manage Realms via RMI calls, while Realms request services from the Normal world through RSI calls.

The life cycle of a Realm always starts from a virtual machine monitor (VMM), \eg qemu~\cite{qemu}, running in the Normal world.
The VMM allocates memory that initially belongs to the Normal world and communicates with the hypervisor, \eg KVM, indicating that it wants to use this memory to create a new Realm.
The hypervisor, through a series of RMI calls, communicates with the RMM.
The ownership of the allocated physical memory granules is passed to the Realm world and the RMM starts building the stage-2 page tables for the Realm.
Realm execution is driven entirely from the Normal world: the Normal-world scheduler schedules the VMM, which transitions into the Realm via an RMI. The Realm itself makes no scheduling decisions and does not require dedicated cores.
As a result, the VMM has full control over the execution of the Realm and can further revoke its memory at any point in time.
When this happens, the RMM guarantees that the memory is first zeroed before returning the ownership to the Normal world, thus preserving confidentiality.

\arch{} CCA provides a built-in attestation mechanism to allow a remote party to verify the state of the Realm and the platform on which it runs before provisioning secrets or granting access.
Each attestation token includes a \textit{Realm Initial Measurement} (RIM), \ie a cryptographic hash of the Realm’s initial code, data, and configuration and serves as its stable identity, and a \textit{Realm Extensible Measurements} (REM).
Realms can update the REM via a dedicated RSI to record measurements of dynamic events during runtime.
Together, RIM and REM allow a verifier to reason about both how the Realm was created and how it reached its current state.

%% file: images/cca-arch.tex
\begin{figure}[t]
    \centering
    \includegraphics[width=1\linewidth]{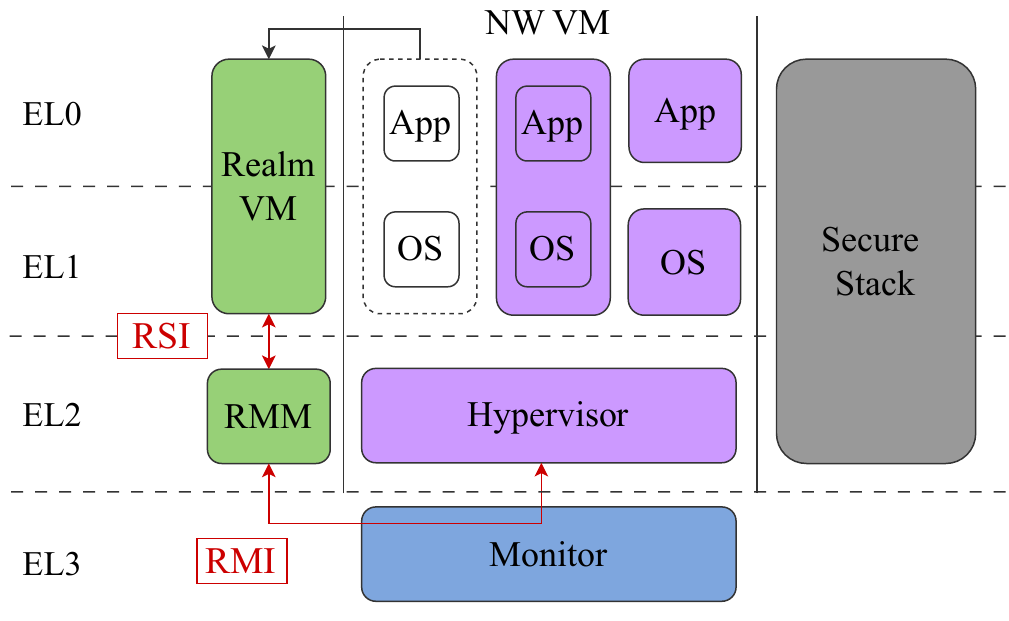}
    \caption{CCA Architecture.}
    \label{fig:cca-arch}
\end{figure}